\documentclass{iaus}
\usepackage{graphicx}
\title[Disk Winds and X-ray Variability] 
{The Physics of Disk Winds, Jets, \\and X-ray Variability in GRS 1915+105}

\author[Neilsen, Lee, \& Remillard]
{Joseph Neilsen$^{1,2},$ Julia C.\ Lee$^{1,2}$, \and Ron Remillard$^{3}$}

\affiliation{$^1$Harvard University Department of
  Astronomy, Cambridge, MA 02138, USA\\[\affilskip]
  $^2$Harvard-Smithsonian Center for Astrophysics, Cambridge, MA
  02138, USA\\[\affilskip] $^{3}$ MIT Kavli Institute for Astrophysics
  and Space Research, Cambridge, MA 02139, USA}
\pubyear{2010}
\volume{275}
\jname{Jets at all Scales}
\editors{Gustavo E.\ Romero, Rashid A.\ Sunyaev, \& Tomaso Belloni, eds.}
\begin{document}

\maketitle

\begin{abstract}
We present new insights about accretion and ejection physics based on
joint \textit{RXTE/ Chandra} HETGS studies of rapid X-ray variability
in GRS 1915+105. For the first time, with fast phase-resolved
spectroscopy of the $\rho$ state, we are able to show that changes in
the broadband X-ray spectrum (\textit{RXTE}) on timescales of seconds
are associated with measurable changes in absorption lines
(\textit{Chandra} HETGS) from the accretion disk wind. Additionally,
we make a direct detection of material evaporating from the
radiation-pressure-dominated inner disk. Our X-ray data thus reveal
the black hole as it ejects a portion of the inner accretion flow and
then drives a wind from the outer disk, all in a bizarre cycle that
lasts fewer than 60 seconds but can repeat for weeks. 
We find that the accretion disk wind may be sufficiently massive to
play an active role in GRS 1915+105, not only in quenching the jet on
long timescales, but also in possibly producing or facilitating
transitions between classes of X-ray variability.\vspace{-2mm}

\end{abstract}\vspace{-7mm}
\section{Introduction}
Of all the known Galactic black holes, GRS 1915+105 is undoubtedly the
most prolific source of state transitions. Discovered as a transient
by GRANAT in 1992 (\cite[Castro-Tirado et al.]{C92}), it has remained
in outburst for the last 18 years and is typically one of the very
brightest sources in the X-ray sky. It is also one of the most
variable: its X-ray lightcurve consists of at least 14 different
patterns of variability, most of which are high-amplitude and
highly-structured (\cite[Belloni et al.\ 2000, Klein-Wolt et
  al.\ 2002, Hannikainen et al.\ 2005]{B00,K02,H05}). It is believed
that many of these variability classes, which are labeled with Greek
letters (\cite[Belloni et al.\ 2000]{B00}), are limit cycles of
accretion and ejection in an unstable disk (\cite[Belloni et
  al.\ 1997, Mirabel et al.\ 1998, Tagger et al.\ 2004]{B97,M98,T04}).

Of its many classes of X-ray variability, three of the best studied are the
$\chi$ state, which produces steady optically thick jets (see,
e.g. \cite[Dhawan et al.\ 2000, Klein-Wolt et al.\ 2002]{D00,K02}),
the $\beta$ state, a wild 30-minute cycle with discrete ejection
events (\cite[Mirabel et al.\ 1998]{M98}), and the $\rho$ state, which
is affectionately known as the ‘heartbeat’ state for the similarity of
its lightcurve to an electrocardiogram (see Figure 1, right panel) and
which consists of a slow rise followed by a short bright pulse,
repeating with a period of roughly 50 s (\cite[Taam, Chen, \& Swank
  1997, Belloni et al.\ 2000]{TCS97,B00}). And while these variability
classes have clearly established strong connections between the
accretion disk and the jet, the physical processes behind this
disk-jet connection have yet to be completely revealed.\vspace{-3mm}

\section{Results and Discussion}
In an effort to quantify the physics of these unusual variability
classes and the disk-jet connection in GRS 1915+105, we recently
analyzed $\sim10$ years of high-resolution \textit{Chandra} HETGS
observations of this black hole X-ray binary (\cite[Neilsen \& Lee
  2009]{NL09}) and performed follow-up with detailed variability
analysis (\cite[Neilsen et al.\ 2010b]{N10b}). In our long-term study,
we found that during states dominated by jet activity,
the X-ray spectra revealed a broad iron emission line, which we
argued must originate when the jet illuminates the inner accretion
disk. In contrast, we found that states where the jet is quenched
display strong, narrow, blueshifted absorption lines from a
highly-ionized accretion disk wind (see also \cite[Miller et
  al.\ 2008]{M08}). We were able to demonstrate that the strength of
this wind is anticorrelated with the \textit{fractional} hard X-ray
flux, which is therefore a useful diagnostic of both the accretion
state and outflow physics. Furthermore, we discovered that the wind
carries enough matter away from the black hole to suppress the jet, so
that mass ejection is regulated in GRS 1915+105 (Figure 1, left
panel).

\begin{figure}[t]
\centerline{
\includegraphics[width=0.45\textwidth]{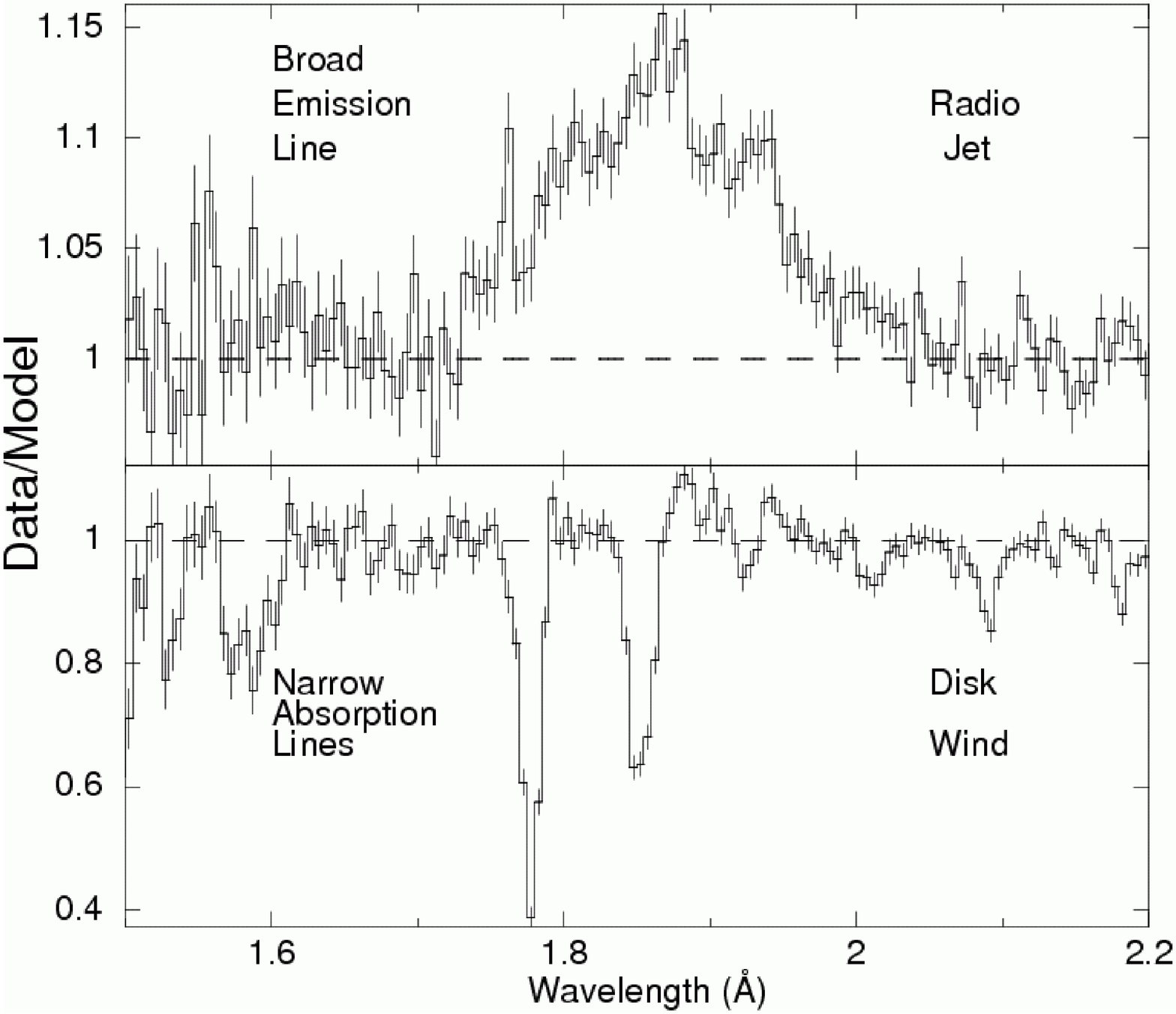}\hfill
\includegraphics[width=0.47\textwidth]{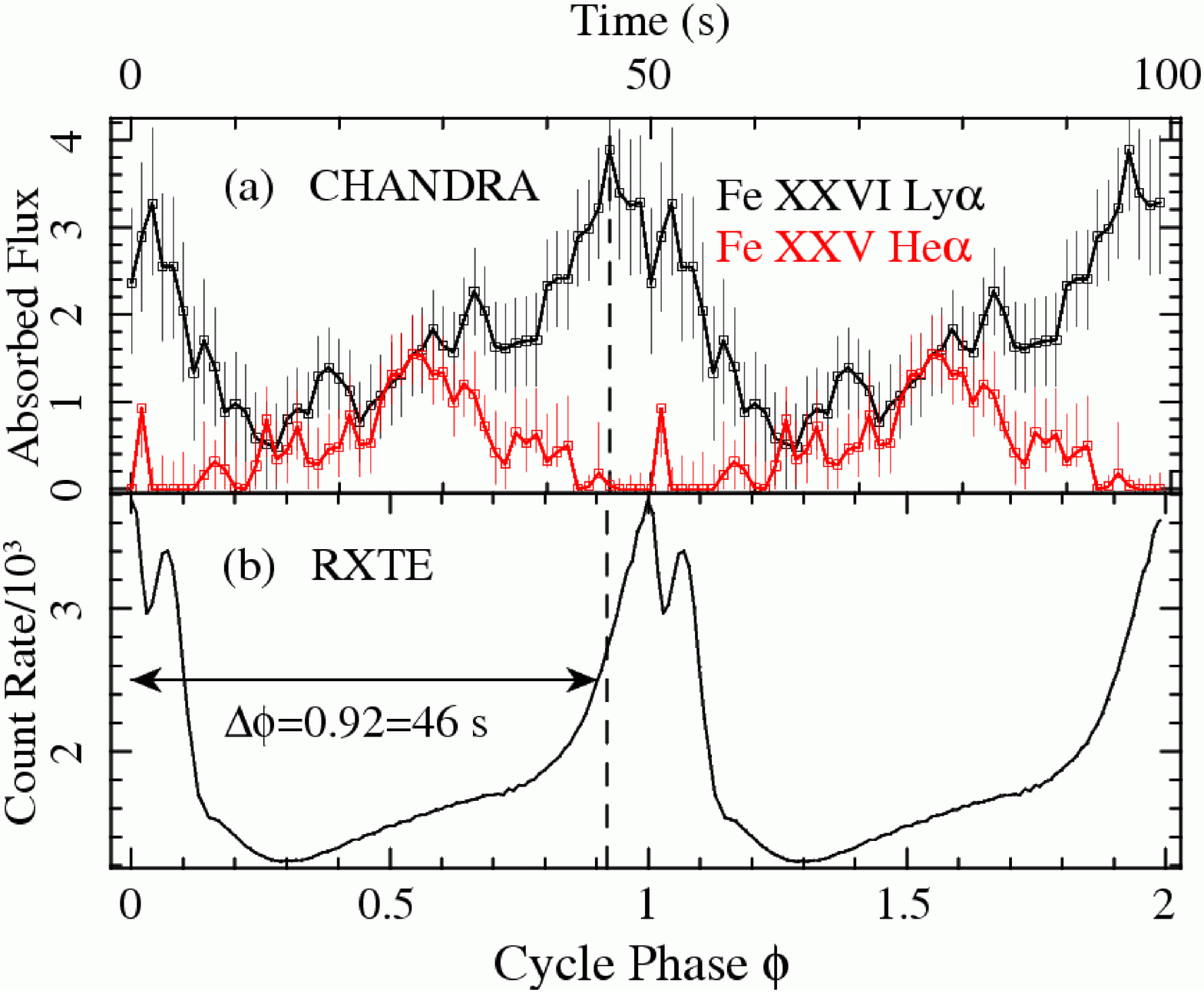}
}
 \caption{Two results from high-resolution X-ray spectroscopy of GRS
   1915+105, clearly indicating links between the disk, wind, and jet
   on many different timescales. \textbf{Left:} The wind-jet
   connection on long timescales, based on \cite[Neilsen \& Lee
     (2009)]{NL09}, where we show that the disk wind may carry enough
   gas away from the black hole to halt the flow of matter into the
   jet. We detect the wind and the jet via their characteristic
   spectral signatures: (top) a broad iron emission line produced when
   the base of the jet illuminates the inner disk, and (bottom)
   narrow, blueshifted absorption lines originating in the
   highly-ionized disk wind. \textbf{Right:} Phase-resolved
   spectroscopy of the accretion disk wind in the $\rho$ state from
   \cite[Neilsen et al.\ (2010b)]{N10b}, showing strong flux and
   ionization changes in the disk wind on timescales of 5 seconds. The
   X-ray lightcurve, phase-folded over many cycles, is shown for
   comparison. Since the X-ray luminosity variations are insufficient
   to produce the observed ionization variability, our phase-resolved
   spectral analysis requires changes in the structure of the wind on
   timescales much less than one minute.}
   \label{fig1}
\end{figure}
To explore the implications of this result on short timescales and
reveal fast (1 s) changes in the accretion flow, we have also
performed the very first phase-resolved spectroscopy of the $\rho$
variability class in GRS 1915+105 (\cite[Neilsen et
  al.\ 2010b]{NRL10}) using a joint \textit{RXTE/Chandra}
observation. Through a combination of X-ray timing and both broadband
and high-resolution X-ray spectroscopy, we show for the first time
that changes in X-ray continuum on timescales of seconds (as probed by
\textit{RXTE}) are associated with measurable changes in absorption
lines (\textit{Chandra} HETGS) from the accretion disk wind (Fig.\ 1,
right panel). Because the X-ray luminosity does not change enough to
produce the observed ionization, the density of the disk wind must be
modulated periodically on these timescales (see \cite[Lee et
  al.\ 2002]{L02} for additional evidence of wind structural changes).

From our broadband spectral analysis, we find
spectroscopic evidence for a \textit{local} Eddington limit
(\cite[Fukue 2004, Lin et al.\ 2009]{F04,LRH09}) and the radiation
pressure instability in the inner accretion disk (e.g.\ \cite[Lightman
  \& Eardley 1974, Belloni et al.\ 1997]{L74,B97}). Our phase-resolved
spectral analysis also allows us to detect bremsstrahlung emission
from material evaporating from the inner accretion flow (see also
\cite[Janiuk \& Czerny  2005]{JC05}). Follow-up comprehensive
analysis of all \textit{RXTE} observations of the $\rho$ state
(\cite[Neilsen et al.\ 2010a]{NRL10}) suggest that this periodic
evaporation process is an essential component of this strange class of
variability.

\section{Conclusion}
In summary, our combined X-ray timing and spectral analysis probing
timescales from 1 second to 10 years reveals new evidence for physical
processes that connect the accretion disk, the radio jet, and the
accretion disk wind. Because the implied mass loss rate in the wind
could be much higher than the accretion rate in the inner disk, we
argue that the wind may be massive enough to play an integral role in
GRS 1915+105, not only in quenching the jet on long timescales, but
also in possibly producing or facilitating transitions between classes
of X-ray variability (\cite[Shields et al.\ 1986, Neilsen \& Lee 2009,
  Luketic et al.\ 2010, Neilsen et al.\ 2010b]{NRL10}).

\end{document}